# Faint Blue Galaxies and the Epoch of Dwarf-Galaxy Formation

ARIF BABUL

Department of Physics, New York University, 4 Washington Place
New York, NY 10003
email: babul@alMuhit.physics.nyu.edu

HENRY C. FERGUSON[1]

Space Telescope Science Institute, 3700 San Martin Drive
Baltimore, MD 21218
email: ferguson@stsci.edu

## ABSTRACT

Several independent lines of reasoning, both theoretical and observational, suggest that the very faint ($B \gtrsim 24$) galaxies seen in deep images of the sky are small low-mass galaxies that experienced a short starburst at redshifts $0.5 \lesssim z \lesssim 1$ and have since faded into low luminosity, low surface brightness objects. We examine this hypothesis in detail in order to determine whether a model incorporating such dwarfs can account for the observed wavelength-dependent number counts as well as redshift, color and size distributions.

Low mass galaxies generically arise in large numbers in hierarchical clustering scenarios with realistic initial conditions. Generally, these galaxies are expected to form at high redshifts. Babul & Rees (1992) have argued the formation epoch of these galaxies is, in fact, delayed until $z \lesssim 1$ due to the photoionization of the gas by the metagalactic UV radiation at high redshifts. We combine these two elements, along with simple heuristic assumptions regarding star formation histories and efficiency, to construct our bursting dwarf model. The slope and the normalization of the mass function of the dwarf galaxies are derived from the initial conditions and are not adjusted to fit the data. We further augment the model with a phenomenological prescription for the formation and evolution of the locally observed population of galaxies (E, S0, Sab, Sbc, and Sdm types). We use spectral synthesis and Monte-Carlo methods to generate realistic model galaxy catalogs for comparison with observations. We find that for reasonable choices of the star formation histories for the dwarf galaxies, the model results are in very good agreement with the results of the deep galaxy surveys. Furthermore, the results of recent gravitational lensing studies, of clustering studies of faint galaxies, of faint galaxy size distributions suggested by analyses of deep HST images, and of regular galaxies at intermediate/high redshifts favor a dwarf-dominated model as an explanation for the faint blue galaxies.

---

[1]Hubble Fellow.

We also discuss various tests of the model based on some generic predictions. For example, the model predicts that the number counts in the $K$-band ought to begin to rise more steeply at magnitudes fainter than $K_{AB} \approx 24$–25. The model also predicts that the local field luminosity function ought to exhibit a steep upturn at magnitudes fainter than $M_B \approx -16$. The detection of the latter, however, depends sensitively on the selection criteria used to construct galaxy catalog. We also consider the possibility of detecting the low surface brightness remnants at low redshifts.

*Subject headings:* galaxies: formation — galaxies:luminosity function, mass function — galaxies:photometry

## 1. Introduction

The faint galaxies seen in deep images of the sky (see Koo 1986; Tyson 1988; Cowie et al. 1988) have been a source of great puzzlement since their detection. Data accumulating over the past decade suggest that the field galaxy population has undergone a dramatic evolution in the past 5–7 Gyrs. In particular, the observations reveal a population of galaxies, first appearing at $B_{AB} \approx 23$ [2] (Tyson 1988; Colless et al. 1993; Koo & Kron 1992), that have colors as blue as those of irregular galaxies today and in some cases, significantly bluer than those of the bluest galaxies found locally, a tendency that is generally accepted as evidence for active star formation. It is this faint blue population that we are most interested in.

As we discuss below, the observed properties of the faint blue galaxy population argue against them being directly associated with the known present-day population of galaxies and suggest that the faint blue galaxies represent a population of galaxies that become visible only while they experience a moderate starburst and thereafter, simply vanish principally either by merging with larger galaxies or fading. For this reason, we shall often refer to these faint blue galaxies as "boojums". (The word "boojums" is an acronym for "blue objects observed just undergoing moderate starburst" and has its origins in the Lewis Carroll's poem "The Hunting of the Snark".)

---

[2] Most magnitudes in this paper are expressed in the $AB$ system (Oke 1974), where $m = -2.5 \log f_\nu - 48.60$. Conversions to the Johnson system are $B = B_{AB} + 0.11$, $I = I_{AB} - 0.48$, and $K = K_{AB} - 1.8$. The wide $B_J$ band used in many deep surveys is approximately $B_{AB} - 0.07$.

The faint blue galaxies (or boojums) pose severe problems for standard models of galaxy evolution and cosmology. First, the total galaxy number-magnitude counts at the faint magnitudes greatly exceed the expectations of standard cosmological models ($0 < q_o \leq 0.5$) containing only a non-evolving population of locally observed galaxies. As illustrated in Figure 1 of Lilly (1993), the galaxy counts are higher by a factor of 4–6 at $B_{AB} \approx 24$ and a factor of 6–8 higher at $B_{AB} \approx 27$ (see also Figure 8). At the same time, the fraction of boojums rises rapidly towards increasingly fainter magnitudes to become a sizable, if not the dominant, fraction (Colless et al. 1993; Koo & Kron 1992).

Second, statistically complete redshift surveys to the limits attainable to date (that is, $B_{AB} \approx 24$ and $I_{AB} \approx 22.5$) reveal that the shape of the redshift distribution of faint galaxies is consistent with no-evolution expectations (Cowie, Songaila, & Hu 1991; Lilly 1993; Colless et al. 1993; Glazebrook et al. 1995); the median redshift of galaxies to $B_{AB} = 24$ is only $z \approx 0.4$ (Cowie, Songaila, & Hu 1991; Glazebrook et al. 1995). For galaxies beyond the current limits of spectroscopy, the fact that the $U$ band observations of the boojums do not show evidence for the Lyman break having been redshifted through the passband (Guhathakurta, Tyson, & Majewski 1990) constraints the redshifts to be $z < 3$. More recently, gravitational lensing studies by Smail, Ellis, & Fitchett (1994) and Kneib et al. (1994) suggest that the bulk of the $B_{AB} \approx 26$–27 galaxies are neither low-$z$ nor high-$z$ galaxies but rather, typically lie at redshifts $z \sim 1$.

Third, deep HST observations reveal that the boojums are very small (Mutz et al. 1994; Im et al. 1995). For a universe with $q_0 = 0.5$, the angular diameter-redshift relation turns over and $\theta(z)$ begin to increase beyond $z \approx 1.2$ (Sandage 1961). While the details are somewhat dependent on $k$-corrections and the galaxy mix, this feature becomes evident in the $\theta(m)$ relation at magnitudes well above the detection limits of HST. Comparison of predictions from standard no-evolution and passive evolution models to the observed angular size distribution from the HST Medium Deep Survey (MDS) reveals a significant discrepancy (Im et al. 1995). Indeed, the angular sizes of galaxies continue to decrease with magnitude with roughly constant slope to the limits of the observations (see Figure 2 of Griffiths et al. 1995). Such observations favor dwarf-dominated models over models involving passive evolution of the large, bright, locally observed galaxy population or a large population of low surface-brightness galaxies (Ferguson & McGaugh 1995).

Finally, the faint blue galaxies have been found to exhibit very weak clustering in comparison to the optically-selected local bright galaxy population (Efstathiou et al. 1991; Neuschaefer, Windhorst, & Dressler 1991; Couch, Jurcevic, & Boyle 1993; Roche et al. 1993; Infante & Pritchet 1995; Brainerd, Smail, & Mould 1995). It is often suggested that the clustering of these objects may be comparable to that exhibited by present-day late type galaxies, if one allows for evolution in the clustering. While this may be true for the brighter of the faint galaxies (that is, galaxies with $B_{AB} \lesssim 24$), studies such as that of Brainerd, Smail, & Mould (1995) show that the clustering properties of the boojums can only be reconciled with this suggestion if either their clustering evolved at an implausibly rapid rate or they are high redshift galaxies. The above discussion about the redshift distribution of the faint blue galaxies argues against the latter possibility.

Early models of faint galaxy counts and redshift distributions started with the observed present-day galaxy mix and worked backwards assuming constant co-moving density of the galaxies, with evolution only in the stellar populations (*e.g.* Tinsley 1980; Yoshii & Takahara 1988; Koo 1990; Guiderdoni & Rocca-Volmerange 1990). In such models, most of the star formation occurs at relatively high redshifts ($z_f \gtrsim 5$), and recent evolution is mild, yielding a redshift distribution that is little different from the no-evolution case, with the exception of a high-redshift tail whose amplitude is extremely sensitive to assumptions about $z_f$, the star formation rate at high $z$, and the amount of dust mixed in with the stars during the starburst epoch. To match the observed optical counts, such models require the cosmology of the universe to be either open or $\Lambda$-dominated so that there is sufficient volume out to intermediate redshifts to account for the large numbers of observed faint galaxies. For $q_0 \lesssim 0.2$, however, such models overpredict faint counts in the K band (Yoshii & Takahara 1988; Yoshii & Peterson 1995). Otherwise, such passive-evolution models successfully reproduce the color-apparent magnitude trends, including the increase in the fraction of blue objects beyond $B_{AB} = 22$ (Koo 1990), and size–magnitude relations (at least to the limits of ground-based resolution). The most serious problem with such models is the rather robust prediction that many galaxies at and beyond $B = 24$ should have redshifts $z \gg 1$, which runs counter to most of the observations.

Gronwall & Koo (1995) have recently considered a variant of the passive-evolution model which starts with the colors, counts, and redshift distributions of faint galaxies and attempts to derive local color distributions and luminosity functions. With $q_0 = 0.05, H_0 = 50$, and $z_f = 5$, the model is reasonably successful, and in particular shows good agreement with the redshift distributions and $K$-band counts, where standard passive-evolution models do not. Apart from the different approach to fitting the data, the two major departures of the model from previous passive-evolution models are (1) inclusion of dust in a way that is not tied to the chemical evolution of the galaxies (as it is, for example, in the models of Guiderdoni & Rocca-Volmerange 1990 or Wang 1991) and (2) the inclusion of a *non-evolving* population of very blue galaxies (their classes 1-3). To the extent that this latter population contributes to the counts, the model in a sense avoids discussing how the faint-blue galaxies came to be and where they are at present. If the bluest classes of the Gronwall & Koo really represent a starburst population, they must either leave a large population of faded remnants, not included in the model, or must have a very high duty cycle of repeated bursts (in which case their colors would be different than assumed). In spite of this criticism, the Gronwall & Koo model is important in demonstrating that the problem of the high-redshift tail can perhaps be solved by a non-standard treatment of dust in normal galaxies and the inclusion of a class of very blue, low-luminosity galaxies.

An alternative class of models does not conserve the co-moving density of galaxies, but rather incorporates *wholesale* merging of the large number of faint galaxies seen at moderate redshifts into large galaxies at the present epoch (Guiderdoni & Rocca-Volmerange 1990; Guiderdoni & Rocca-Volmerange 1991; Broadhurst, Ellis, & Glazebrook 1992). In these models, the mass of a typical galaxy mass grows with time such that a typical $L_*$ galaxy is approximately a factor 2–3 less massive at redshift $z = 0.5$ and a factor 3–5 less massive at $z = 1$. There are various problems with such scenarios. First, theoretical and $N$-body studies of realistic models for hierarchical structure formation (Lacey & Cole 1993; Kauffmann, White, & Guiderdoni 1993; Lacey & Cole 1994; Navarro, Frenk, & White 1994) show that high merger rates required by the above models are unlikely within the framework of a dynamical friction-driven merging scheme. Second, the present-day spiral galaxies in the heuristic *wholesale* merging models are predicted to accrete as much as 40–60% of their mass over the past 5 Gyr; Tóth & Ostriker (1992), however, argue that a typical spiral disk cannot have accreted more 10% of its mass in the past 5 Gyr without the disks being thicker than is observed. If the faint galaxies preferentially merged into present-day elliptical galaxies, then there ought to be more blue light in the latter than is observed unless the faint galaxies experienced rapid fading (Dalcanton 1993). If fading is invoked, then the merging hypothesis becomes superfluous as fading itself can account for the steep decline in the galaxy counts. Third, if the boojums are declining in numbers due to wholesale mergers, one would expect that those at $B_{AB} \gtrsim 24$ ought to be at least as strongly clustered as the local population of bright galaxies. As discussed previously, most studies seem to indicate otherwise. (It should be noted that the clustering results do not argue against occasional mergers, only against *wholesale* mergers.) Finally and most importantly, observational studies of galaxies selected on basis of their giving rise to MgII absorption lines in spectra of background QSOs suggest that the present-day population of bright galaxies (both early-type as well as late-type) was already well-established by $z \sim 1$ and has undergone no significant evolution since then in luminosity, color, size or space density (Steidel & Dickinson 1994; Steidel, Dickinson, & Persson 1994). The authors are led to conclude that there must be "two distinct galaxy populations [present at $z \lesssim 1$], one of which must be evolving very rapidly while the other remains stable" and that "evolution of the faint blue galaxies apparently goes largely unnoticed by the [bright] galaxies".

Associating the boojums with a population of dwarf galaxies undergoing a short burst of star formation at moderate redshift that have since faded away is the basis of an another class of models (Lacey & Silk 1991; Babul & Rees 1992; Gardner, Cowie, & Wainscoat 1993; Lacey et al. 1993; Kauffmann, Guiderdoni, & White 1994). The main differences among these models are the mechanisms for triggering and/or suppressing star formation in the dwarf galaxies. In the Lacey et al. (1993) model, the triggering mechanism is tidal interactions with normal galaxies. Kauffmann, Guiderdoni, & White (1994) propose no specific triggering mechanism, but their "bursting CDM" also assumes that dwarf galaxies only form stars in proximity to a larger galaxy. While the low-mass halos in such models are expected to be only weakly clustered (Efstathiou 1995), the fact that those that form stars are clustered much like regular galaxies leads to the expectation that the angular correlation function of boojums ought to similar to that expected for the regular bright galaxies (if not even stronger if each bright galaxy has a cluster of bursting dwarfs in its vicinity). However, the observed correlation amplitude for the boojums is very small, suggesting that these galaxies are very weakly clustered (for example, see Brainerd, Smail, & Mould 1995). In the Babul & Rees (1992; hereafter BR92) model, on the other hand, the increase in the starburst activity in low-mass halos is correlated with the decline in the UV background in the universe, which enables the gas in the low-mass halos to cool, collapse and undergo star formation. In this case, the likelihood of star formation in a given halo is not governed by the proximity of neighboring galaxies, although there may be some modulation by quasars and AGN through the proximity effect.

In this paper we adopt the principal idea of the BR92 model, namely that the boojums are dwarf galaxies whose formation has been delayed until $z \lesssim 1$ by photoionization. The mass function and size distribution of the dwarf galaxies are derived from fairly general considerations and are not adjusted to fit the data. An important feature of the model is that the number density of dwarf-galaxy halos is *not* a free parameter. Hence the $N(m)$ and $N(z)$ relations for these dwarf galaxies depends only on their star formation histories and the selection functions of the surveys. Using Monte Carlo simulations of deep surveys, we show that with plausible choices for the star formation histories of the dwarf galaxies and their expansion during the super-nova wind phase, as well as with a plausible model for the evolution of the locally observed population of galaxies (E, S0, Sab, Sbc and Sdm types), we can achieve reasonable agreement with the observations. In subsequent papers, we will compare the models in detail to deep, high-resolution observations from HST (Ferguson, Giavalisco, & Babul 1995).

In §2 of this paper, we outline and expand upon the BR92 model. In §3, we describe our passive evolution model for the locally observed population of galaxies, as well as the spectral-synthesis and the Monte-Carlo methods used to simulate deep surveys. In §4, we discuss the evolution of the individual galaxy types in our model and of the collective properties of the galaxy population. We compute, and compare with observations, the galaxy number counts as a function of magnitude in different bands, the redshift distribution, the luminosity function (local), the color distribution, and the distribution of effective galaxy radii. The pros and the cons of our model, and some possible tests, are discussed in §5. Finally, we present a summary in §6. Throughout this paper, we assume the $\Omega = 1$ Einstein-de Sitter model for the universe and adopt a Hubble constant of $H_0 = 50 h_{50}$ km s$^{-1}$ Mpc$^{-1}$.

## 2. Boojums as Starbursting Dwarf Galaxies

In keeping with the model outlined by BR92, we adopt the generally accepted hierarchical clustering scenario for galaxy formation as outlined by White & Rees (1978). In such scenarios, galaxies arise when density perturbations in a universe dominated by an unspecified type of dissipationless dark matter collapse under the action of gravity to form virialized dark halos, with star formation ensuing when the gaseous component of the halos cools and condenses in the central cores. The mass and the size (virial radius) of dark halos, in terms of redshift and the halos' circular velocity $V_c = \sqrt{GM(R)/R}$, are

$$M = 7.5 \times 10^9 h_{50}^{-1} V_{35}^3 \left(\frac{1+z}{2}\right)^{-3/2} M_\odot,  \tag{1}$$

where $V_{35} = V_c/35$ km s$^{-1}$ and

$$R = 13.3 h_{50}^{-2/3} \left(\frac{M}{10^9 M_\odot}\right)^{1/3} \left(\frac{1+z}{2}\right)^{-1} \text{ kpc}.  \tag{2}$$

For simplicity, we assume that the virialized dark halos have spherically symmetric density profiles

$$\rho(r) \propto 1/(r^2 + r_c^2),  \tag{3}$$

where $r_c$ is a core radius. In absence of any natural length scale for the core radius, we assume that $r_c \propto R$ and more specifically,

$$r_c = 1.0 h_{50}^{-2/3} \left(\frac{M}{10^9 M_\odot}\right)^{1/3} \left(\frac{1+z}{2}\right)^{-1} \text{ kpc}.  \tag{4}$$

It should be noted that our adopted value for the core size is not inappropriate; local gas rich dwarf galaxies, whose rotation curves have been well-studied and which appear to be almost completely dominated by dark matter, have density profiles that can be approximated as isothermal spheres with core radii of order 3–7 kpc (see Moore (1994) and references therein).

BR92 associate the faint blue galaxies with minihalos having $15 \lesssim V_c \lesssim 35 \, \mathrm{km \, s^{-1}}$. In generic hierarchical models for galaxy formation, such minihalos should condense from the expanding background and virialize at high redshifts. For example, in the standard $\Omega = 1$ cold dark matter (CDM) model for structure formation with a power spectrum normalized such that the rms density fluctuation in a sphere of radius $16 h_{50}^{-1} \, \mathrm{Mpc}$ is $\sigma_8 = 0.67$, the bulk of $M \sim 10^9 \, M_\odot$ halos virialize at $z \approx 3.5$. Ordinarily, one would expect that the minihalos would experience starburst soon after virialization. However, observations of Ly$\alpha$ clouds at high redshifts suggest at these redshifts, the universe is permeated by a metagalactic UV flux originating in quasars and young galaxies. At $z = 2$, this background is estimated to have a strength $J_\nu \approx 10^{-21} \, \mathrm{erg \, cm^{-2} \, s^{-1} \, Hz^{-1}}$ at the Lyman limit (Bechtold et al. 1987). Such an ionizing flux is intense enough to photoionize the gas in collapsing minihalos and maintain it at a temperature of $T \approx 3 \times 10^4 \, \mathrm{K}$. In dark halos with $15 \, \mathrm{km \, s^{-1}} \lesssim V_c \lesssim 35 \, \mathrm{km \, s^{-1}}$, the photoionized gas would be stably confined, neither able to escape nor able to settle to the center (Rees 1986; Ikeuchi 1986), until the intensity of the metagalactic ionizing flux declines sufficiently.

Although the epoch-dependence of the UV flux is uncertain at higher redshifts, it seems almost certain that at smaller redshifts, $J_\nu$ is a steeply declining function of $z$. The sources of the UV background, whether they be quasars or young galaxies, are likely to be less intense per comoving volume at low redshifts. In the absence of any sources at $z < 2$, $J_\nu$ would scale as $(1 + z)^{3 + \alpha}$, where $\alpha$ is the spectral index. A more detailed analysis by (Zuo & Phinney 1993) suggests that the $J_\nu$ declines as $(1 + z)^3$.

A declining $J_\nu$ has two effects on the gravitationally bound gas cloud in the minihalos. First, the equilibrium temperature of the highly photoionized gas depends, albeit somewhat weakly, on $J_\nu$ (see Black 1981) and this would cause a gradual concentration towards the center (Ikeuchi, Murakami, & Rees 1989). Second, the gas becomes more neutral and offers a greater optical depth to the ionizing photons, leading to a diminution of the ionizing flux reaching the central regions and the formation of a warm ($T \approx 9000 \, \mathrm{K}$) shielded neutral core (Murakami & Ikeuchi 1990) in pressure equilibrium with an ionized envelope. The formation of a neutral core, however, is not a sufficient condition for star formation to ensue. For a gas cloud to be susceptible to star formation, it must be at least marginally self-gravitating (e.g. Matthews 1972). To satisfy this constraint, the entire baryonic component of a minihalo must accumulate within a central region of size

$$R_b \approx 1.9 h_{50}^{-2/3} \left( \frac{M}{10^9 \, M_\odot} \right)^{1/3} \left( \frac{1+z}{2} \right)^{-1} \text{kpc} \tag{5}$$

if $\Omega_b = 0.1$, and this can occur only if the gas in the neutral core can cool below 9000 K.

Further cooling of a neutral metal-poor gas is only possible if $H_2$ molecules form. The formation of molecular hydrogen in a gas of primordial composition occurs via gas phase reactions (see Shapiro & Kang 1987). In the presence of ionizing radiation, molecular hydrogen will not form efficiently until the region in question is effectively shielded against photons with energies greater than 25 eV that photoionize helium (Kang et al. 1990). BR92 have argued that the rapid decline in the intensity of the background UV flux at $z < 2$, coupled with the increasing efficiency in shielding of the central regions, will give rise to appropriate conditions for the formation of molecular hydrogen at $z \sim 1$ (see also Babul & Rees 1993). Thereafter, the temperature of the gas in the core plummets down to $\sim 10^2$ K. With the resulting loss of pressure support from the core, the baryons in the ionized envelope will also collapse (the phenomenon is analogous to that studied by Shu 1977) and accumulate at the center of the potential well.

The foregoing argument suggests that most of the low-mass halos, despite having virialized early, would experience a long "latency" period before forming stars. During this period, some of the stable minihalos may merge and become incorporated into larger systems (and therefore, no longer be stable minihalos) while new minihalos may condense out of the general background. In hierarchical clustering scenarios with realistic initial conditions on galactic and sub-galactic scales (i.e. spectral index $-2 \lesssim n \lesssim -1$), the distribution of low mass halos is a steep function of mass ($d \ln N / d \ln M \approx -2$). For example, assuming that the power spectrum of density fluctuations that grow into halos of interest is adequately described by the standard CDM power spectrum normalized such that $\sigma_8 = 0.67$, that is

$$\sigma(M) = 7.5 \left( \frac{M}{10^9 \, M_\odot} \right)^{-0.1}, \tag{6}$$

the comoving number density of minihalos existing at $z \approx 1$ estimated using the analytic Press-Schecter (1974) formalism is

$$\frac{dN}{d \ln(M)} \approx 2.3 \left( \frac{M}{10^9 \, M_\odot} \right)^{-0.9} \text{Mpc}^{-3}. \tag{7}$$

According to Lacey & Cole (1993), minihalos existing at $z \approx 1$ have long survival times. Almost all of the minihalos existing at $z \approx 1$ will survive for roughly 5 Gyr and up to 40% will survive to the present.

As originally noted by BR92, the actual redshift at which the UV background ceases to able to prevent baryonic collapse (and the subsequent star formation activity) in a particular minihalo depends sensitively on its particular density profile.[3] It is not possible at present to make a detailed prediction of the redshift range over which this baryonic collapse occurs or its dependence on galaxy mass. For present purposes, we assume that the probability that a given halo will experience starburst at some Hubble time $t$ is described by an exponential decay:

$$p(t) \, dt \propto e^{-t/t_*} \, dt, \tag{8}$$

where $t_*$ is exponential decay timescale, the formation timescale for the dwarf population as a whole. We further assume that no halo will experience a starburst at redshifts $z > 1$. With such a burst probability, we can, by varying the value of $t_*$, explore the consequences of having a nearly constant probability from $z = 1$ to the present (large values of $t_*$) or the consequences of having all the halos bursting at redshifts close to 1 (small values of $t_*$).

Within an individual halo, the natural scaling for star formation rate in a self-gravitating gas cloud is

$$\dot{M}_* \propto \frac{M_g(t)}{t_{ff}}, \tag{9}$$

where $M_g(t)$ is the instantaneous mass of gas cloud and $t_{ff}$ is the free-fall timescale for the cloud. As noted by BR92, the duration of starburst in the minihalos is expected to be relatively short ($\tau_{SB} \approx 10^7$ yr); the very first generation of supernovae will expel all the gas out of the dwarf galaxy, quenching further star formation (see also Larson 1974; Dekel & Silk 1986). We therefore assume that during starburst, the star formation rate is constant and equal to the initial value.

For a gas cloud of mass $M_g = \Omega_b M$ and radius $R_b$ at the center of a minihalo of mass $M \sim 10^9 \, M_\odot$, the star formation rate is of order unity and is proportional to mass. We express the star formation rate as

$$\dot{M}_* = \epsilon_* \frac{M}{\tau_{SB}} = 100 \epsilon_* \left( \frac{M}{10^9 \, M_\odot} \right) \left( \frac{\tau_{SB}}{10^7 \, \mathrm{yr}} \right)^{-1} M_\odot \, \mathrm{yr}^{-1}, \tag{10}$$

where $0 < \epsilon_* \leq \Omega_b$ is the fraction of the *total* halo mass that is converted into stars by the end of the starburst. In effect, $\epsilon_*$ represents the star formation efficiency and characterizes our ignorance regarding details of the supernovae heating of the interstellar medium and of the mass loss.

---

[3]Chiba & Nath (1994) argue that the decline in $J_\nu$ at low redshifts is unlikely to trigger the formation of low-mass galaxies. However, we believe this is a result of their using top-hat density profiles for low mass halos. Consider two objects of same size $R$ and mass $M = 4\pi\rho_0 R^3$, one being a singular sphere with density profile $\rho(r) = \rho_0 (R/r)^2$ and the other a uniform sphere with $\rho(r) = 3\rho_0$. The column density (optical depth) through the center is finite for one and infinite for the other. A top-hat model is the limiting case of a density profile that would be *least* able to cool in the presence of metagalactic ionizing background radiation.

In a starburst galaxy, the flux at wavelengths shortward of the Balmer discontinuity in a starburst galaxy is dominated by massive young stars and therefore, the UV luminosity of the galaxy is primarily determined by the instantaneous star formation rate (White & Frenk 1991). Furthermore, the observed B-band flux from a galaxy lying in the redshift range $0.3 < z < 3$ is a direct measure of this luminosity. Consequently, a galaxy at redshift $z$ with a star formation rate $\dot{M}_*$ will be observed to have a B-band magnitude

$$B_{AB} = -13.17 - 2.5 \log(l_B/1 \, \mathrm{erg \, s^{-1} \, cm^{-2}}), \tag{11}$$

where,

$$l_B = L_{UV}(\dot{M}_*/1 \, M_\odot \, \mathrm{yr^{-1}})\Delta\nu(1+z)/4\pi d_L^2, \tag{12}$$

and $\Delta\nu = 1.6 \times 10^{14} \, \mathrm{Hz}$ is the effective bandwidth, $d_L = 1.2 \times 10^4 h_{50}^{-1}(1+z)[1-(1+z)^{-1/2}] \, \mathrm{Mpc}$ is the luminosity distance, and $L_{UV} \approx 6.7 \times 10^{26} \, \mathrm{erg \, s^{-1} \, Hz^{-1}}$ is the UV luminosity of the star-forming region assuming a Salpeter IMF truncated at $0.1 \, M_\odot$ and $100 \, M_\odot$. A galaxy of mass $10^9 \, M_\odot$ that converts 3% of its total mass (30% of its baryonic mass if $\Omega_b = 0.1$) into stars during a $\tau_{SB} = 10^7 \, \mathrm{yr}$ burst at $z = 0.4$ would have an apparent magnitude of $B_{AB} \approx 24.9$. At $z = 1$, the same galaxy would have an apparent magnitude o $B_{AB} \approx 26.7$. For a fixed value of $\epsilon_*$, the contribution of the dwarfs to faint galaxy counts is clearly sensitive to the duration of the starburst. If the star formation timescale were more than a few times $10^7$ years the dwarfs would not be bright enough to contribute to the counts near $z = 1$. On the other hand, very short burst phases would produce too many galaxies brighter than $B = 24$ unless the star formation efficiency were much lower as well. Thus, in order to dominate the faint galaxy counts, the dwarfs at $z \approx 1$ must form stars within a timescale that fortuitously turns out to be roughly that expected if supernovae terminate the starburst.

The expulsion of the gas from the galaxy, and the resulting reduced gravity, causes the remnant stellar system to expand. The degree of expansion, however, is moderated by the presence of the dark halo. If the timescale for mass loss is longer than the dynamical timescale in the central star forming region, the stars in the galaxy respond adiabatically to the mass loss. The radii of the orbits before and after the mass are related by $r_i M_i(r_i) \approx r_f M_f(r_f)$, where $M(r)$ is the total mass (baryons + dark matter) within radius $r$, and the subscripts $i$ and $f$ refer to the initial and the final states respectively (Dekel & Silk 1986; Lacey et al. 1993). In a galaxy that converts 30% of its baryons ($\Omega_b = 0.1$) into stars and expels the rest, the stellar system will expand a factor of 1.3. The degree of expansion under the assumption that the mass loss occurs instantaneously is comparable.

The key assumption of our model is that star formation occurs in bursts. If star formation in the low-mass halos predicted by CDM is quiescent, extending over more than a few times $10^7$ yr, then these galaxies will never be bright enough to influence the faint galaxy counts. There is ample evidence from studies of nearby dwarf galaxies that star formation does occur in bursts, and that galactic winds play a key role in influencing the star-formation histories (e.g. Searle, Sargent, & Bagnuolo 1973; Meurer et al. 1992; Marlowe et al. 1995; Armus et al. 1995; Tosi 1993). Nevertheless, the model described above is clearly a vast oversimplification of the real world. In general, dwarf galaxies will not be spherical, will not all have the same density profile, will not have a step-function star-formation history, and will not expel all their remaining gas at the end of the starburst. Furthermore the stellar populations (described in §3) will undoubtedly not all have the same initial mass function and metallicity. An ideal physical model would predict the distribution functions of all of these parameters and integrate over the population to compare to deep survey observations. However, this is well beyond current capabilities. Instead, we assume that in a broad statistical sense all of these details can be subsumed in the parameters $t_*$, which governs the probability distribution of star-formation episodes as a function of redshift, and $\epsilon_*$, which governs the luminosity released in an average star-formation episode. The simplifying assumption of a single burst of star-formation followed by ejection of all the remaining gas is thus not essential to our model, as long as the star-formation occurs in episodes with a relatively high star formation rate (1 to 10 $M_\odot \, \mathrm{yr}^{-1}$) with an overall time-evolution of the burst probability similar to our assumed exponential law. The essential features of our model (see §4) — that the population at $B > 25$ is primarily low-mass galaxies at $z < 1$, that the population is weakly clustered, and that the remnant population still exists at present as low-surface brightness dwarfs — are largely independent of the details of the star-bursts within the individual galaxies.

## 3. Testing the Bursting Dwarf Hypothesis

The simplest and the most efficient way of comparing the model to observations is through the analysis of model galaxy catalogs, generated by Monte Carlo methods, that incorporate the appropriate survey selection criteria and magnitude estimation schemes. The Babul-Rees (or the "boojums as bursting dwarfs") hypothesis, in and of itself, only specifies the formation and evolution of low-mass "dwarf" galaxies. In order to generate realistic galaxy catalogs for comparison with observations, we combine the fading dwarfs model with a prescription for the formation and evolution of the locally observed population of galaxies. We divide this latter population into five galaxy types: E, S0, Sab, Sbc and Sdm. Our model is thus a hybrid approach that works forwards from initial conditions for the dwarfs, but works backwards from present day properties for the "normal" galaxies. The rationale for this approach is that the star formation histories along the traditional Hubble sequence are reasonably well constrained from observations of nearby galaxies, but not very well constrained from hierarchical structure models. On the other hand, the steep mass function of the dwarf galaxies is *robust* consequence of hierarchical structure built from CDM-like power spectra, even if such low mass galaxies are not well represented in the *observed* local luminosity function.

The procedures for creating the model catalogs can be summarized as follows: Galaxy candidates and their internal properties are drawn at random from distribution functions for galaxy type, total luminosity, bulge/total mass ratio (our galaxies are made up of bulges and disks whose stellar populations and luminosity profiles are modeled separately), surface brightness, redshift and age. For each galaxy considered, we follow the evolution of its stellar population to the required age and compute the associated synthetic spectrum. The galaxy is then "observed" and included in the catalog if it conforms with the selection criteria. These steps and our specific assumptions are described in more detail in the following subsections.

## 3.1. Evolution of the Stellar Population

### 3.1.1. Model Inputs

In order to compute the synthetic spectra of each galaxy under consideration, it is necessary to trace the evolution of its stellar population as a function of the galaxy's age. Stellar evolution theory provides us with the effective temperature $T_{\rm eff}(M, Z, t)$ and bolometric luminosity $L_{\rm bol}(M, Z, t)$ of a star of mass $M$ amd metallicity $Z$ as a function of time. Ignoring chemical evolution, one can determine the stellar population of a galaxy given a complete set of $L_{bol}$ and $T_{eff}$ evolutionary tracks as well as the initial mass function $\phi(M)$, which specifies the number of stars of mass $M$ born per unit mass when a stellar population is born, and the star formation rate $\Psi(t)$. To compute accurate spectra at early times requires interpolation between evolutionary tracks for stars of different masses, a procedure that is equivalent to computing isochrones. Rather than start with the evolutionary tracks, therefore, it is more convenient to begin with isochrones which give $L_{\rm bol}, T_{\rm eff}$ on a fine grid of $M, t$, as advocated by Charlot & Bruzual (1991) and Bruzual & Charlot (1993).

Due to the paucity of theoretical models covering all phases of stellar evolution, Charlot & Bruzual (1991) were forced to piece together isochrones from a variety of theoretical evolutionary tracks and empirical sources (for AGB stars). Since the Bruzual & Charlot (1993) models, new isochrones have become available that cover a wider range of stellar masses and phases of evolution in a self-consistent manner (Bertelli et al. 1994). Hence, we avoid some of the difficulties encountered by Charlot & Bruzual (1991) in trying to merge tracks computed by different authors with different sets of opacities. The new isochrones have been used for spectral synthesis by Bressan, Chiosi, & Fagotto (1994), to which the reader is referred to for a summary of the basic physical assumptions underlying the stellar evolution models. Our galaxy models are constructed from stellar populations of a single metallicity. For E, S0, Sab, Sbc, and Sdm galaxies, we use $Z = 0.02$ (*i.e.* $Z = Z_\odot$) for both the bulge and the disk stars. For the dwarfs, we use $Z = 0.001$ ($Z = Z_\odot/20$; [Fe/H] $= -1.3$). One of the advantages of the Bertelli *et al.* isochrones is that the isochrones cover a wide range of metallicities: $Z = 0.0004, 0.001, 0.004, 0.008, 0.02$, and $0.05$, with corresponding helium abundances of $Y = 0.23, 0.23, 0.24, 0.25, 0.28$ and $0.352$. We are therefore in a position to construct more realistic spectra for low-metallicity galaxies than previously possible for models of faint galaxy counts.

We adopt an exponential star formation law for the bulges and disks of the non-dwarf galaxies, and tune their star formation histories so that their present-day colors (for the total bulge+disk light) match "standard" colors of the local galaxies. Table 1 shows the adopted colors, and the corresponding $e-$folding timescales for star formation in the bulges and disks of the different galaxy types. For the dwarfs, we assume a starburst with a constant star formation rate over $1 \times 10^7$ yr. Thereafter, the galaxy undergoes pure passive evolution (no star formation).

### 3.1.2. Computing Model Spectra

We begin by assuming that stars form according to the Salpeter initial mass function, $dN/dM \propto M^{-2.35}$, with lower and upper mass limits of 0.1 and $100 M_\odot$. Using the isochrone synthesis algorithm, we derive theoretical values for the bolometric luminosity $L_{\rm bol}$ and effective temperature $T_{\rm eff}$ of the stars in a population as a function of initial mass and age. For each group of stars of a given initial mass and age, we next search our catalog of synthetic stellar spectra (based entirely on model atmospheres of Kurucz 1992 for $T_{\rm eff} < 5000$ K and of Clegg & Middlemass 1987 for $5000 < T_{\rm eff} < 200000$ K) for the nearest match in $T_{\rm eff}$, and then (at fixed $T_{\rm eff}$) in $\log(g)$. The gravity of a star is

$$\log(g) = \log(M/M_\odot) + 4\log(T_{\rm eff}) - log(L/L_\odot) - 10.604, \tag{13}$$

where $M$ is the *current* mass of the star, taking into account the mass lost in all previous stages of evolution, as tabulated by Bertelli et al. (1994).

Fig. 1.—
**FIGURE AND CAPTION GOES HERE**

The spectrum of the galaxy is the weighted sum of the different model atmosphere spectra. The weight accorded to the flux from each model atmosphere is:

$$w = 4\pi r^2 T_c g_c \phi(M) dM \Psi(t) dt,$$ (14)

where

$$r = \sqrt{GM/g}$$ (15)

is the radius of the star, the initial mass function $\phi(M)$ specifies the fraction of the total number of stars formed at any given time with masses in the range $M$ and $M + dM$, and the star formation rate $\Psi(t)$ specifies the fraction of the population being of a particular age. The remaining factor $(T_c g_c)$, where

$$T_c = T_{\text{eff}}^4 / T_{\text{eff}}^4(\text{grid}) \quad \text{and} \quad g_c = g(\text{grid})/g,$$ (16)

corrects for the mismatch between the actual temperatures and gravities of the stars, and those available in the grid. The factor ensures that the bolometric flux is correct even though detailed spectral shape may be slightly off due to the finite $T_{\text{eff}}$, $\log(g)$ sampling of the model atmosphere grid.

Our model spectra do not include emission lines from HII regions or supernova remnants, but do include extinction due to dust during the rapid star formation phase of giant galaxies. The addition of dust is motivated by the success of the models of Wang (1991) and Gronwall & Koo (1995) in removing the high-redshift tail from standard passive-evolution models. While the amount of dust and the shape of extinction curve in principle must depend on chemical evolution, in practice there is only a weak dependence on metallicity seen in the dust properties of nearby starburst galaxies (Calzetti, Kinney, & Storchi-Bergmann 1994; Storchi-Bergmann, Calzetti, & Kinney 1994). The form of the extinction probably depends mostly on how the stars and dust are mixed together.

We have included dust in our models only during the rapid formation phases of ellipticals, bulges and S0 disks. Dust will *inevitably* accompany star formation in such systems, but modeling its evolution is beyond the scope of this paper. For our purposes, we have simply adopted the "extinction curve" of Calzetti, Kinney, & Storchi-Bergmann (1994), with $\tau_B$ varying smoothly from 0 at the birth of the galaxy, to $\tau_B = 1$ at the peak of the star formation epoch, falling to $\tau_B = 0$ again by late epochs, as shown in Figure 1. While the model is ad-hoc, these optical depths are probably underestimates. Nevertheless, they are sufficient to remove high-redshift galaxies from samples down to $B = 24$. For our purposes, then, the details of the dust evolution do not matter so long as the attenuation is $\tau_B \gtrsim 1$ during the starburst phase.

Fig. 2.—
**FIGURE AND CAPTION GOES HERE**

Fig. 3.—
**FIGURE AND CAPTION GOES HERE**

Spectra of E and S0 galaxies have been set to zero beyond the Lyman limit.

### 3.1.3. Uncertainties in the Models

It is well-known that synthetic spectra from model atmosphere do not provide a perfect match to real stellar spectra. The most serious discrepancies arise at the extremes of the temperature and gravity ranges sampled by the model atmosphere grids. At low temperatures, molecular opacity becomes important but is not fully included in the model atmospheres. At high temperatures and low gravities, non-LTE effects become important.

A standard procedure for circumventing these difficulties is to use empirical spectra of individual stars, especially at cool temperatures. This is the technique adopted, at least in part, by Bruzual & Charlot (1993) and Bressan, Chiosi, & Fagotto (1994). For simplicity and expediency, we have chosen to rely entirely on model atmospheres, recognizing that this may cause problems at some ages and wavelengths. The biggest drawback of this approach is the lack theoretical synthetic spectra for stars cooler than $T_{\text{eff}} = 3500$ K. We associate such stars with the coolest Kurucz model atmosphere. Our procedure for computing the model-atmosphere weights ensures that the total bolometric luminosity of the star will be correct; however the flux will be distributed differently from the overall spectral energy distribution of a real M star.

We can estimate the uncertainties associated with our procedure for computing model galaxy spectra by comparing spectra for simple, single-metallicity coeval stellar populations (SSP's) to those computed by Bressan, Chiosi, & Fagotto (1994). The spectra are constructed from the same underlying isochrones, although by different codes. However, Bressan, Chiosi, & Fagotto (1994) supplemented the Kurucz model grid with observed spectra of stars cooler than $T_{\text{eff}} = 3500$ K. Figure 2 shows the comparison for a solar metallicity SSP at ages $10^7, 10^8, 10^9$, and $10^{10}$ years. The differences between the spectra are minor, typically a few hundredths of a magnitude. Figure 3 shows the same comparison for a population of 1/50 solar metallicity. Once again the differences introduced by our reliance on model atmospheres are very minor.

Fig. 4.—

**FIGURE AND CAPTION GOES HERE**

More serious differences are introduced by the use of different theoretical isochrones. The dashed curves in Figure 2 show SSP spectra computed using the Bruzual & Charlot (1993) code, which relies on a different set of underlying isochrones (*e.g.* based on the evolutionary tracks of Schaller et al. 1992 for massive stars). The differences in colors computed from our synthetic spectra and those of Bruzual & Charlot can be up to about 1 magnitude in rest-frame $B - K$, but are typically much smaller.

Since our models are based entirely on theoretical isochrones and theoretical model atmospheres, it is quite simple to investigate the effects of varying the chemical composition. Figure 4 shows our computed spectra for metallicities $Z_\odot$ and $Z_\odot/50$. Changing metallicities introduces differences of up to 3 magnitudes in rest-frame $B - K$ colors at fixed age. *These differences are almost entirely due differences in the isochrones;* model spectra constructed using low metallicity ($Z_\odot/50$) isochrones and solar-metallicity atmospheres differ by at most a few tenths of a magnitude from the curves shown in Figure 4.

We now summarize and rank the importance of the errors introduced by our various approximations. For computing counts and redshift distribution, the most important and least constrained aspect of the model is the early photometric evolution of early type galaxies and spiral-galaxy bulges. As outlined above, our assumptions for the photometric evolution, and in particular the amount of dust, during the star formation epoch in these galaxies, were motivated by the need to remove the high-$z$ tail from the redshift distribution. The star formation rates and amounts of dust needed to do this are entirely plausible. Of course there are other ways to remove the high-redshift tail (*e.g.* by forming all bulges and early type galaxies at low redshift), but we find the local evidence that stars in these systems are mostly old too compelling to adopt this point of view.

Next in importance is any potential mismatch between the metallicity assumed in our models and those of real galaxies. Our assumption of single metallicity is a vast oversimplification, but is difficult to overcome without constructing a chemical evolution model for each galaxy type. Generally, we expect that the normal galaxies will be bluer at early epochs than we have assumed. In the case of the dwarf galaxies, our choice of metallicity may have some effect on optical-infrared colors. We do not expect that adding more realistic chemical evolution would significantly alter the predicted counts or redshift distributions because these are primarily determined by the assumptions for the star formation rate vs. time and the initial mass function.

Modifications of the *slope* of the IMF are likely have a very small effect on the slope of the number counts or the median of the redshift distribution (Kauffmann, Guiderdoni, & White 1994). Modifications in the upper and lower mass limits of the IMF do affect the counts, but essentially trade off against the star-formation efficiency $\epsilon_*$. If the lower-mass limit were raised from $0.1 M_\odot$ to $0.5 M_\odot$, the required star-formation efficiency would be significantly lower. Uncertainties in the underlying stellar isochrones probably have a negligible effect on the optical counts and colors, but may influence the $K$-band counts and optical-infrared colors at a level of up to a magnitude. Finally, the adoption of model atmospheres for cool stars instead of empirical stellar spectra probably does not introduce any significant errors in any of the faint-galaxy distribution functions.

## 3.2. Distribution Functions for Locally Observed Galaxies

Apart from being able to follow the spectral evolution of galaxies, we need to specify the mass functions, bulge/disk ratios, and surface brightness distributions for the different galaxy types in order to generate realistic galaxy catalogs.

For the locally observed galaxy population, we normalize the spectra such that galaxies with fiducial "mass" $M_0 = 1.0$ have $M_{B_J} = -21.1$ at an age of $1.3 \times 10^{10}$ years. All luminosity functions (LF's) are then specified in terms of fiducial mass. This automatically corrects for any mismatch between our adopted mass limits for the IMF and the true ones in galaxies.

The luminosity functions for the different galaxy types are Gaussian, with parameters chosen such that the individual LFs approximate those shown in Binggeli, Sandage, & Tammann (1988) while the overall luminosity function has a Schechter (1976) function profile over the range: $M_{B_J} = -23$ to $-16$, with $\phi_*$ set to $2.3 \times 10^{-3}$ gal Mpc$^{-3}$ (Yoshii & Takahara 1988). We note that this assumed luminosity function for locally observed, "normal" population of galaxies has a pronounced deficit of galaxies fainter than $M_{B_J} \approx -15$, which will be filled in by the brighter members of the faded dwarf population.

The Gaussian approximation for the luminosity functions of individual galaxy types implies that the fiducial mass of a given galaxy of a particular type in the Monte-Carlo simulation can be assigned according to:

$$M_0 = 10^{\log <M_0> -0.4 \sigma g},$$
(17)

where $g$ is a Gaussian random deviate with unit variance. The parameters $<M_0>$ and $\sigma$, the mean and the dispersion of the mass functions, are listed in Table 2. In addition, this table also lists the local space density in galaxies per cubic Mpc, integrated over the mass function from a lower limit of $0.0035 <M_0>$ to an upper limit of $10 <M_0>$.

Table 2 also lists our adopted mean bulge-to-total light ratios $(B/T)$ and scatter, taken from Simien & de Vaucouleurs (1986), for the different galaxy types. For a given galaxy, chosen at random from the mass function for a given morphological type, the bulge mass is assigned according to

$$M_0(\text{bulge}) = (B/T)M_0 + g\sigma_{B/T},  \tag{18}$$

where $g$ is a Gaussian random deviate with unit variance, as above. Our calibration of the fiducial mass ensures that the adopted $B/T$ refers to the $B_J$ band at redshift $z = 0$. The spectral dependence and evolution of $B/T$ is automatically accounted for by the separate synthesis models for the bulge and disk components.

We adopt standard relations for the sizes of the locally observed galaxies. The Monte-Carlo procedure allows us to introduce realistic scatter about the mean relations. For the bulges, we use the relation of Sandage & Perelmuter (1990). Because the total magnitude is evolving, we express the relation in terms of fiducial mass. Hence, the absolute magnitude that a galaxy *would* have, if it had an age of $1.3 \times 10^{10}$ years, is

$$X = -2.5 \log M_0 - 21.1  \tag{19}$$

and the corresponding effective surface brightness is

$$Y = (-0.48X + 11.02)(1 + g).  \tag{20}$$

As before, $g$ is a Gaussian random deviate. The corresponding effective radius, for an $r^{1/4}$ law, is

$$r_e = 4.85 \times 10^{-8} \sqrt{10^{(Y-X-0.75)/2.5}/\pi} \text{ kpc}.  \tag{21}$$

For the disks, we use Freeman value for the central surface brightness, with a 0.4 magnitude scatter,

$$X = -2.5 \log M_0 - 21.1  \tag{22}$$

$$Y = 21.6 + 0.4g.  \tag{23}$$

The corresponding exponential scale length for the disk is

$$\alpha = 4.85 \times 10^{-8} \sqrt{10^{(Y-X)/2.5}/2\pi} \text{ kpc}.  \tag{24}$$

All simulated galaxies associated with the locally observed population are assumed to form at $z = 5$ and their redshifts are assigned in a manner that ensures a constant co-moving density.

Table 1: Giant Galaxy Colors and Star Formation Timescales

| Type | Bulge/Tot | Observed Means | | | Model Colors | | | $e-$folding times | |
|------|-----------|---------------|---|---|--------------|---|---|-----------|------|
| | | $U-B$[a] | $B-V$[a] | $B_J-K$[b] | $U-B$ | $B-V$ | $B_J-K$ | bulge | disk |
| E | 1.0 | 0.54 | 0.96 | 3.91 | 0.61 | 1.07 | 3.90 | 0.01 | – |
| S0 | 0.4 | 0.54 | 0.96 | 3.91 | 0.61 | 1.07 | 3.90 | 0.01 | 1.0 |
| Sab | 0.3 | 0.10 | 0.79 | 3.08 | 0.12 | 0.73 | 3.36 | 0.01 | 30.0 |
| Sbc | 0.15 | -0.09 | 0.64 | 3.06 | -0.01 | 0.63 | 3.21 | 0.01 | 30.0 |
| Sdm | 0.0 | -0.15 | 0.54 | 3.06 | -0.14 | 0.50 | 2.83 | – | 30.0 |

[a]From Tinsley (1978).

[b]From Mobasher, Ellis & Sharples (1986).

Table 2: Giant Galaxy Parameters

| Type | Luminosity Function | | | Bulge/Total Ratio | |
|------|---------------------|---|---|-------------------|---|
| | $N_0$ | $<M>$ | $\sigma$ | mean | $\sigma$ |
| E | 0.37 | 0.11 | 1.7 | 1.0 | 0.0 |
| S0 | 1.15 | 0.11 | 1.7 | 0.4 | 0.25 |
| Sab | 2.00 | 0.35 | 1.1 | 0.3 | 0.27 |
| Sbc | 4.00 | 0.09 | 1.3 | 0.15 | 0.27 |
| Sdm | 8.00 | 0.0091 | 1.3 | 0.0 | 0.0 |

### 3.3. Distribution Functions for Dwarf Galaxies

The properties of the dwarf population have already been discussed in §2. The salient features are:

1. The luminosity function for the dwarfs is a power law, given by equation (7), with lower and upper mass cutoffs of $6 \times 10^8 \, M_\odot$ and $7.5 \times 10^9 \, M_\odot$ respectively. (The masses correspond to $V_c = 15$ and $35 \, \mathrm{km \, s^{-1}}$ at $z = 1$.)

2. The dwarf galaxies begin to form at $z = 1$ and experience a short starburst of $10^7$ yrs. From equation (8), the fraction of dwarfs that have not experienced starburst in $\delta t$ years after the beginning of the starburst epoch is

$$f(\mathrm{dwarfs}) = e^{-\delta t/t_*}. \tag{25}$$

3. According to equations (10) and (12), the flux from a dwarf galaxy of mass $M$ at redshift $z$ is

$$l(M) = l_0(M/M_0), \tag{26}$$

where $l_0$ is the flux from a galaxy of mass $M_0 = 1 \times 10^9 \, M_\odot$ at the same reshift.

Within the framework outlined above, the essential free parameters are the star formation efficiency $\epsilon_*$ and the decay rate of the dwarf galaxy formation probability $1/t_*$. Less free, but still somewhat important is the lower limit of the dwarf-galaxy mass function, which is determined to within factors of a few by the ability of $10^4$ K gas to collapse into low velocity-dispersion halos (Efstathiou 1992). Parameters for the different runs are shown in Table 3. $N_0$ in the table is the local space density of dwarfs integrated over the mass function.

We model the surface brightness distribution of the dwarf galaxies with an exponential profile whose scale length is derived from the collapse calculations described in §2.

Table 3: Dwarf Galaxy Parameters

| Model | $M_{\mathrm{lower}}$ ($10^8 M_\odot$) | $t^*$ ($10^8$ yr) | $\epsilon$ | $N_0$ Mpc$^{-3}$ |
|---|---|---|---|---|
| 1 | 6 | 20.0 | 0.03 | 4.0 |
| 2 | 6 | 50.0 | 0.03 | 4.0 |
| 3 | 6 | 1.0 | 0.03 | 4.0 |

Associating half-light radius with the half-mass (baryonic) radius and noting that the stellar systems in the dwarfs will expand during the epoch of mass loss, we set

$$r_e = 1.0(M/M_0)^{1/3} \text{ kpc for galaxy age t} < 1 \times 10^7 \text{ yr} \tag{27}$$

and

$$r_e = 1.3(M/M_0)^{1/3} \text{ kpc for galaxy age t} > 1 \times 10^7 \text{ yr}. \tag{28}$$

We have modeled the expansion as a step function.

### 3.4.   Constructing catalogs of model galaxies

We construct a separate model-galaxy catalog for each of the galaxy types under consideration, selecting the galaxy parameters at random from the distribution functions described above. The procedure has been extensively tested to ensure that the input distributions are properly reproduced, even in the tails of the distributions. These catalogs list the redshifts, masses, ages, the bulge $r_e$ and disk $\alpha$ (in kpc and arcsec), and magnitudes in various bands (for bulge and disk components, as well as the total) for each of the galaxies. The magnitudes are computed by integrating the properly redshifted and $k$-broadened spectra for the bulge and disk components through the filter bandpasses given in Bruzual (1981). The associated zeropoints are given in Ferguson & McGaugh (1995).

These separate catalogs are then processed by different programs that "observe" the galaxies using a given seeing and selection criteria. While observational results are usually quoted in terms of total magnitude, in practice galaxy detection is usually based on flux within a certain area above a certain threshold. To facilitate comparison with observations, we convolve the galaxies in our simulations with a Gaussian seeing profile with the FWHM quoted in the observational papers, then select galaxies and measure magnitudes in a way that closely matches the techniques in the surveys. The details of object selection and photometry in faint galaxy surveys are difficult to model precisely and sometimes are not completely specified in published reports. For those observational surveys that we have chosen to compare to, we list the associated selection criteria in Table 4 (see Ferguson & McGaugh 1995 for more details).

The galaxies that pass the selection criteria are then used to compute the luminosity function, the number-magnitude counts $N(M)$, the redshift distribution $N(Z)$, etc. Galaxies of different morphological types are combined together by weighting the contribution of each galaxy by $N_0(\text{type})/N_{\text{catalog}}(\text{type})$, where $N_{\text{catalog}}(\text{type})$ is the density of galaxies in the input catalog. For normal galaxies, $N_{\text{catalog}}(\text{type})$, is computed using all the galaxies in

Table 4: Assumed Survey Parameters

| Observer | Band | Magnitude | Seeing ($''$) | $D_{min}$ ($''$) | $\mu_{lim}$ | |
|----------|------|-----------|-----------|--------------|-----------|---|
| Loveday | $B_J$ | Isophotal | 3.0 | – | 24.5 | 1 |
| Tyson | $B_J$ | Isophotal | 1.7 | – | 28.7 | |
| Colless | $B_J$ | Isophotal | 1.0 | – | 26.5 | 2 |
| Lilly N(m) | $I_{AB}$ | Hybrid | 1.2 | 2.0 | 28.0 | 3 |
| Lilly N(z) | $I_{AB}$ | Aperture | 0.7 | 3.0 | 28.0 | 4 |
| Cowie K-band N(m) | $K_{AB}$ | Aperture | 1.0 | 3.0 | 25.4 | 5 |
| Djorgovski K-band N(m) | $K_{AB}$ | Aperture | 0.75 | 1.5 | 25.8 | 6 |
| Songaila K-band N(z) | $K_{AB}$ | Aperture | 0.6 | 3.5 | 28.63 | 7 |
| Glazebrook | $B_{AB}$ | Isophotal | 2.26 | 1.6 | 26.6 | 8 |

[1]Correction of -0.27 added to isophotal magnitudes.

[2]Correction of -0.58 added to isophotal magnitudes.

[3]Isophotal magnitudes for galaxies with $D_{28} > 2''$; otherwise, used $2''$ aperture.

[4]Assumed complete to $D_{28} > 2''$. $3''$ aperture used.

[5]Assumed complete to $D_{25.4} > 1''$. Isophotal magnitudes used.

[6]Selection is roughly by total magnitude, as the details are not specified. Magnitudes are within a $1.5''$ diameter aperture, with a constant of -0.275 added.

[7] we use magnitude limits $19.8 < K_{AB} < 20.8$ (Songaila et al. 1994 use $K$ rather than $K_{AB}$), and measure magnitudes in an aperture of $1.75''$ radius. We have not attempted to match the Songaila et al. isophotal selection criteria, as they varied from field to field (see Cowie et al. 1994).

[8]While galaxies are selected by isophotal diameter, magnitudes are computed in a $4.0''$ diameter aperture with a constant -0.3 magnitude correction to total magnitude (this varied slightly from night to night in the real survey).

Fig. 5.—
**FIGURE AND CAPTION GOES HERE**

Fig. 6.—
**FIGURE AND CAPTION GOES HERE**

the volume out to $z = 5$. For the dwarfs, $N_{\text{catalog}}(\text{type})$ is computed from the number of galaxies with $z < 0.1$, to allow for the fact that the space density is evolving.

Our simulations do not explicitly include noise that is present in real observations. To the extent that the algorithms used to detect galaxies in deep surveys are unbiased, the effect of noise is simply to increase the scatter in the measured magnitudes. For the purpose at hand, a scatter of few tenths of a magnitude is unimportant. In any case, the best way to assess the impact of noise would be to construct simulated images with noise and analyze them in the same way as the observations. Such an analysis will be presented in a separate paper in which we perform a detailed comparison of the model predictions with results from deep, high-resolution observations from HST and the NTT (Ferguson, Giavalisco, & Babul 1995).

## 4. Results

### 4.1. Evolution of a Starburst Dwarf Galaxy

Figure 5 shows the evolution of the spectral energy distribution for a $10^9\,M_\odot$ dwarf that experiences a $10^7$ yr starburst, during with time it forms stars of metallicity $Z = Z_\odot/20$ at the rate of $3\,M_\odot\,\text{yr}^{-1}$. During the starburst, the galaxy spectrum is quite steep ($f_\nu \propto \nu$), with the UV light being largely due to short-lived massive stars. Once the starburst ceases and the number of massive stars declines, so does the UV luminosity. The spectral energy distribution becomes increasingly dominated by longer-lived but nonetheless aging, less massive stars. Figure 6 illustrates this more clearly. The top panel shows the evolution of the rest frame $M_{B_j}$ (solid curve) and $M_{K_{AB}}$ (dashed curve) as a function of time. After the cessation of star formation, the B-band luminosity of the galaxy fades by as much as 6 magnitudes over a span of $10^{10}$ yr. At longer wavelengths, the fading is more tempered. The lower panel of Figure 6 shows the time evolution of the dwarf galaxy's rest frame $B_J - K_{AB}$ and $B_J - I_{AB}$ colors. Based on these color curves we can anticipate the conclusion of the next section that the bursting dwarfs, which dominate the counts at $B > 25$ will not contribute significantly to the $I$ and $K$-band counts until near the limits of current surveys.

Fig. 7.—
**FIGURE AND CAPTION GOES HERE**

Fig. 8.—
**FIGURE AND CAPTION GOES HERE**

## 4.2. Evolution of Normal Galaxies

Figure 7 shows the evolution of the rest-frame $B_J$ and $K_{AB}$ magnitudes for E, S0, Sab, Sbc, and Sdm galaxies with present day luminosity $M_{BJ} = -21$ and bulge/disk ratios at the means for each type. Because spiral galaxies in our models consist of bulge and disk components drawn from rather broad distributions in relative luminosity, in practice the normal galaxies in our models fill in the whole range of colors from E to Sdm galaxies. At early times $t < 1$ Gyr, the photometric evolution is dominated by the bulge component, while at late times colors are primarily governed by the bulge/disk ratio. Obviously, the evolution in luminosity shown in Figure 7 includes no merging. However, as long as *most* of the merging that formed ellipticals and bulges occurred above redshifts $z \approx 2$ (Kauffmann 1995), the photometric evolution will be essentially identical by $z = 1$ (the limits of current redshift surveys). At $z \approx 1$, the distribution of $\mathcal{R} - K$ colors in our models is consistent with that observed in samples of galaxies selected by virtue of Mg II absorption in the spectra of QSO's (Steidel, Dickinson, & Persson 1994; Steidel & Dickinson 1994). At lower redshifts, the luminosity-evolution of all galaxy types in our model is modest. With respect to the present-day luminosity, the galaxies at $z = 1$ are 0.1 to 0.7 mag brighter in rest-frame $B_J$. In the $K$-band, the evolution ranges from $-0.3$ to 0.6 mag.

## 4.3. The Number-Magnitude Counts

Having described the evolution of individual galaxy types, we now turn to the collective properties of the complete galaxy sample. We begin by comparing the number-magnitude trends predicted by the model with observations. In Figure 8, we plot the data for the number counts in $B_J$ ($\approx B_{AB} - 0.07$), $I_{AB}$, and $K_{AB}$ bands drawn from various sources, as well as the number-magnitude relationship for a non-evolving population of locally observed galaxies (hereafter, referred to as the NE model). Comparing the observed $B$-band counts versus the NE model, one finds that by $B_{AB} = 24$ the NE curve falls short by a factor of four. The $N(M)$ relations for our models, computed using total magnitude selection, are shown for different values of $t_*$, the formation timescale for the dwarf population. Bearing in mind that curves are computed on the basis of total magnitudes and that the incorporation of selection effects only acts to decrease the number of galaxies observed at a given magnitude, it is evident that formation timescales as short as $10^8$ yr (*i.e.* most of the dwarf galaxies form very soon after $z \approx 1$) are not favored. This conclusion will be further strengthened when we consider the corresponding redshift distribution. It is, however, important to point out that all the models predict that the $K$ counts ought to begin to rise somewhat more steeply at magnitudes fainter than $K_{AB} \approx 24$–25. This is a generic feature of the bursting dwarf model.

Fig. 9.—
**FIGURE AND CAPTION GOES HERE**

Fig. 10.—
**FIGURE AND CAPTION GOES HERE**

Choosing Model 1 ($t_* = 2 \times 10^9$ yr) as our fiducial model, we compute the $N(m)$ in the different bands according to the selection criteria summarized in Table 4. The results are shown in Figure 9. The model number counts (solid curve) are in very good agreement with observations in all three bands. For comparison, we also plot the number counts computed with total magnitude selection, as well as the NE curve. The effect of selection on these models is to flatten the observed N(m) relation at faint magnitudes near the limits of ground-based surveys. In Figure 10, we decompose the number counts associated with Model 1 (total magnitude selection) into counts due to the evolving, locally observed galaxy population (dotted curve) and the starbursting dwarf population (dashed curve). In the B-band, the excess (with respect to NE model) of galaxies in the range $18 \lesssim B_{AB} \lesssim 23$ is largely due to the evolving population of locally observed galaxies. Thereafter, the starbursting dwarfs become increasingly important. 30% of the galaxies $B_{AB} \leq 24$ are starbursting dwarfs and the fraction increases towards fainter magnitudes. The same is true for the excess in the I-band ($18 \lesssim I_{AB} \lesssim 23$). The fraction of dwarfs in a $I_{AB} \leq 23$ sample is approximately 15%. It should be noted that the dwarf fraction does not take into account the contribution of other small but normal evolving galaxies such as the Sdms.

### 4.4. The Redshift Distribution

From the above discussion it is clear that the dwarf galaxies in these models do not dominate the counts at the limits probed by current redshift surveys. *The bursting dwarf galaxy model cannot by itself resolve the discrepancy between the $N(z)$ and $N(m)$ relations in the range $20 < B < 24$.* Thus comparison to the redshift distributions only weakly test dwarf-galaxy aspect of the model. Mostly, the comparisons test our treatment of selection biases and the modifications we have made to the standard passive evolution model (the inclusion of separate bulge and disk components with a distribution function in fractional contribution to the total light, and the incorporation of dust at early epochs).

Fig. 11.—
**FIGURE AND CAPTION GOES HERE**

In Figure 11, we compare the predicted redshift distribution to the observations for three values of $t_*$, where the galaxies were subject to the selection criteria of the Glazebrook et al. (1995) survey. The model distributions are normalized so that the number of galaxies in the redshift range $0 \leq z \leq 2$ matches than in the survey. This comparison provides our primary constraint on $t_*$. Models with a prolonged epoch of dwarf galaxy formation ($t_* \approx 5 \times 10^9$ years) overpredict the number of low redshift galaxies that would be detected at $B = 24$ even with the appropriate selection effects taken into account. On the other hand, models with a very short epoch of formation ($t_* \approx 10^8$ yrs), predict a pronounced peak in the redshift distribution at the redshift of formation of the dwarfs. The model with a moderate formation timescale, $t_* = 2 \times 10^9$ yr, yields a redshift distribution that is in reasonable agreement with the observations.

While the peak of the redshift distributions in the models and the data agree very well, in all cases the models distributions appear somewhat flatter than the observed distribution. As the Glazebrook et al. (1995) redshift survey is only 73% complete, inconsistency at the level shown in Figure 11 is not a problem, and indeed is *expected* if the unidentified galaxies tend to lie at higher redshifts. Glazebrook et al. (1995) discuss the possible redshift distribution of the unidentified sources (sources brighter than the $B = 24$ magnitude limit that were detected but whose redshifts were not obtained). They conclude that the unidentified sources are most likely to be a mixture of weak-lined sources in the redshift range $0 < z < 1$, and galaxies with $z > 1$, for which [OII] is shifted out of the spectral window. The color distribution of the unidentified objects favors the interpretation that they are mostly late type galaxies at $z > 0.5$.

In Figure 12, we present the redshift distributions associated with our fiducial model (Model 1: $t_* = 2 \times 10^9$ yr), computed using different selection criteria. Selection criteria and magnitude estimation schemes resemble those used in the real surveys (Colless et al. 1993; Lilly 1993; Glazebrook et al. 1995; Songaila et al. 1994), and the models are normalized to the data over the redshift range shown. The most complete surveys, those by Colless et al. (1993) and Lilly et al. (1995) show good agreement with the models, while the deeper, less complete surveys show distributions more sharply peaked at $z \sim 0.5$ than the models, which may be an artifact stemming from the difficulty of obtaining redshifts for galaxies at higher redshift.

Fig. 12.—
The $N(z)$ redshift distribution for our fiducial ($t^* = 2$ Gyr) model, compared to the data from various surveys. The models have been normalized to match the total number of objects observed in the redshift range $0 < z < 2$. The number of galaxies predicted with $z > 2$ ranges from 4 to 12%, being the highest for the Lilly (1995) selection criteria. The $K$-band redshift sample consists of galaxies observed in fields SSA4, SSA13, and SSA17, and SSA22, by Songaila et al. (1994), with the number of galaxies normalized to account for variations in area coverage over the magnitude range $18 < K < 19$ (hence there are non-integral numbers of galaxies in the plot).

We conclude by emphasizing that the redshift distributions primarily test our adopted models for the evolution of normal galaxies. Table 5 shows the relative contributions of different galaxy types to the redshift distributions shown in Fig. 12. Bursting dwarf galaxies in the model contribute significantly only to the Glazebrook et al. (1995) survey. Given the incompleteness, the only distribution that appears problematical is the $K$-band sample of Songaila et al. (1994). However that sample contains only 24 galaxies, so a definitive test will have to await the results of larger surveys.

Why have previous passive evolution models for normal galaxies predicted a high-redshift tail? The attempt to match the counts at very faint magnitudes ($B > 25$) forces such models to have the galaxies be visible during the star formation phase (i.e. not be hidden by dust), and to have a rather prolonged epoch of formation (so as not to produce a feature in the $N(m)$ distribution). While such a galaxy formation scenario is plausible (see for example Guiderdoni & Rocca-Volmerange 1990) it is by no means a unique way to arrive at the colors of present-day normal galaxies. By providing an alternative explanation for the very faint galaxies, our model removes the requirement that normal galaxies, particularly the giant galaxies, be visible during their formation phase. For a redshift of formation $z_f \approx 5$, it is galaxies with star formation timescales $\tau \approx 1 \times 10^9$ years that are most problematical for the redshift distributions. Such galaxies do not fade sufficiently by redshifts of 1-3 to disappear from the redshift distribution. By including dust during the star formation epoch of ellipticals, bulges, and S0 disks, we have quite consciously *tuned* the models to remove the high redshift tail. The required optical depths are modest (peaking at $\tau = 1$ in the rest-frame B band). While other explanations for the lack of high-redshift galaxies in the existing deep surveys are clearly possible, we contend that existing redshift surveys do not require serious modifications to passive evolution models, or to the idea that the normal galaxy population formed largely at redshifts significantly greater than unity.

Table 5: Percent contribution of different morphological types to $N(z)$

| Survey | Band | E+S0 | Sab+Sbc | Sdm | Dwarf |
|---|---|---|---|---|---|
| Colless | $B_J$ | 14 | 67 | 7 | 12 |
| Lilly | $I$ | 29 | 65 | 5 | 1 |
| Glazebrook | $B$ | 14 | 53 | 6 | 27 |
| Songaila | $K$ | 41 | 57 | 2 | 0 |

Fig. 13.—
**FIGURE AND CAPTION GOES HERE**

Fig. 14.—
**FIGURE AND CAPTION GOES HERE**

For illustrative purposes, we plot the redshift distribution of galaxies brighter than $B_J = 26$ for our fiducial model in Figure 13. We simply selected the galaxies on the basis of total magnitudes since there are no redshift surveys that reach such faint magnitudes. Three features of the redshift distribution are noteworthy. First, the low redshift peak is due to relatively nearby, fading dwarf remnants. As long as the galaxies are selected on the basis of total magnitudes, this peak is present even if the threshold magnitude is raised to $B_J = 24$. Real surveys, however, tend to be biased against low surface brightness objects and therefore cannot detect these remnants. (Note the lack of a low redshift peak in Figure 12.) Second, the tail at $z > 1$ is entirely due to passively evolving, locally observed population of galaxies. Third, the peak at $z \approx 1$ is due to starbursting dwarf population. This peak sharply cuts off at $z = 1$ because of our fiat that dwarf halos form stars only at $z \leq 1$.

### 4.5.   Luminosity Function

In Figure 14, we plot the overall present-day luminosity function for galaxies in Model 1, selected by total magnitude as well as by isophotal selection criteria of Loveday et al. (1992) survey. We also plot the luminosity functions for each morphological type. As noted previously (§3.2), the luminosity functions for the different galaxy types, with the exception of the dwarfs, are Gaussian, with parameters chosen such that the individual

Fig. 15.—
**FIGURE AND CAPTION GOES HERE**

LFs approximate those shown in Binggeli, Sandage, & Tammann (1988) while the overall luminosity function has a Schechter (1976) function profile over the range: $M_{B_J} = -23$ to $-16.5$, the limit of the Loveday et al. (1992) survey for $H_0 = 50$ km s$^{-1}$ Mpc. Over this magnitude range, the isophotal selection yields similar results to total magnitude selection.

The presence of dwarfs in Model 1 causes the total magnitude luminosity function to steepen at magnitudes fainter than $M_{B_J} = -16$. There is some evidence for such a rise in the local galaxy luminosity function. Ferguson & Sandage (1991) found that the luminosity function in nearby loose groups turns up below $M_B = -16$, with the faint end being dominated by low-surface brightness dE galaxies. More recently, the luminosity function of galaxies in the $m_Z \leq 15.5$ cfA Redshift Survey (Marzke, Huchra, & Geller 1994) also shows an an indication of such a rise, as does the luminosity function in rich clusters of galaxies (De Propis et al. 1995). The ability to detect the predicted turnup in the luminosity function depends sensitively on the selection criteria of the survey. For example, the luminosity function recovered from our simulated catalogs using the isophotal selection criteria of Loveday et al. (1992) survey shows no evidence of the steepening in spite of the large number of low luminosity dwarfs present in Model 1.

While their space density is high, the luminosity density for the faded dwarfs is significantly lower than for the normal galaxies. Adding up the light from the galaxies in the simulated catalogs used to produce Figure 14, we arrive at estimates in the $B_J$ band of $1.1 \times 10^8 L_\odot$ Mpc$^{-3}$ for the normal galaxies and $3.3 \times 10^7 L_\odot$ Mpc$^{-3}$ for the dwarfs. In the $K$ band, the numbers are $4.9 \times 10^8$ and $1.7 \times 10^7$, respectively. Thus the dwarfs constitute at most 3% of the luminosity density of the local universe.

Looking back towards $z \approx 1$, the "regular" galaxies experience very modest luminosity evolution, with galaxies being brighter by 0.1 to 0.7 mag in rest-frame $B_J$ and -0.3 to 0.6 mag brighter in $K$-band.

## 4.6. Color Distributions

The comparison of the observed versus model color distributions provides an additional test of the model. In Figure 15, we compare the Model 1 galaxy colors to those observed by Lilly, Cowie, & Gardner (1991). Applying their selection criteria, the galaxies were required to have $20 < I_{AB} < 25$ and $B_{AB} < 27$ through a $3''$ aperture in a $1.2''$ seeing. The mean

Fig. 16.—
**FIGURE AND CAPTION GOES HERE**

Fig. 17.—
**FIGURE AND CAPTION GOES HERE**

Fig. 18.—
**FIGURE AND CAPTION GOES HERE**

$B_{AB} - I_{AB}$ color is 1.51 in the data and 1.26 in the models, with most of the difference lying in the tails of the distribution. The peak of the color distribution agrees to within 0.2 mag. The model galaxy sample consists of 27% dwarfs. Hence, the color distribution at these magnitude limits primarily tests the assumptions we have made for the evolution of normal galaxies. At fainter magnitudes, the colors become rapidly bluer as the dwarf galaxies begin to dominate the numbers. This trend is very much in keeping with that illustrated by Figure 2 of Koo & Kron (1992), which shows that the color distribution of galaxies with $B_J \lesssim 23$ is similar to that of the nearby galaxies while at fainter magnitudes, the distribution shifts significantly towards the blue.

In Figure 16, we plot $I - K$ (Johnson) color versus K-magnitude for Model 1 galaxies selected at random from the simulated catalogs with $I_{AB} < 25.48$ with a sample size equal to that of Djorgovski et al. (1995). This is the equivalent of their Figure 4, and is broadly consistent, although perhaps lacking somewhat in blue galaxies with $K < 21$. We particularly draw attention to the fact that the models do not predict an overabundance of very blue galaxies at the faint limits, in spite of the fact that the dwarfs are predominantly very blue.

### 4.7. Effective Radii

For $q_0 = 0.5$, the angular-size *vs.* redshift relation has a minimum at $z \approx 1.2$. The counts at the limits of ground based surveys in our models are dominated by dwarf galaxies with effective radii $\sim 1$ kpc at redshifts $0.5 \lesssim z \lesssim 1$. At these redshifts the angular effective radii of the dwarf galaxy population are a few tenths of an arcsecond. The angular size distribution predicted by our models probably comes close to the extreme that could be expected for *any* model with a geometrically flat universe. While we reserve detailed size comparisons to a later paper (Ferguson, Giavalisco, & Babul 1995), we show in Figure 17 a comparison to recent measurements from the Medium Deep Survey (MDS) (Griffiths et al.

1995) in the magnitude range $21 < I < 22$. If anything, the models contain *too few* galaxies with effective radii less than 0.5 arcsec. We have not attempted to duplicate the MDS selection criteria and it is possible the selection criteria may tend to favor the detection of galaxies with small effective radii due to their higher surface brightnesses. In any case, at $I = 22$, the models are still dominated by normal galaxies. The predicted distribution at fainter magnitudes is shown in Figure 18.

## 5. Discussion

The tests in the previous section show that the "boojums as starbursting dwarfs" model satisfies the available constraints from ground-based measurements of the number counts, redshift distributions, and colors of faint galaxies, and shows qualitative consistency with early HST measurements of angular sizes. The success is admittedly due partly to parameters that are not well constrained. For the normal galaxies, the ad-hoc inclusion of dust during the star formation epoch cures the tendency for passive-evolution models to overpredict the number of high-$z$ galaxies in deep redshift surveys. For the dwarfs, the parameter $t_*$ hides our ignorance of the detailed evolution of the ionizing background and the distribution of gas density profiles in proto-dwarf galaxy halos. To match the counts and redshift distributions, $t_*$ must be approximately $2 \pm 1 \times 10^9$ yrs. Having to a certain extent adjusted the models to fit the data, the agreement of the model color and size distributions with the data provides some confirmation that our assumptions are correct. However, the tests performed in this paper do not prove that bursting dwarf galaxies are the correct solution to the faint blue galaxy problem, merely that the hypothesis is not ruled out.

### 5.1. Pros and Cons of the model

From a theoretical perspective there are several attractive features to the bursting dwarf model. First, the mass function and space density of dwarf galaxy halos are *not* free parameters, having been fixed by the adoption of a realistic power spectrum such as the CDM power spectrum to describe the initial density fluctuations on galactic and sub-galactic scales. Second, the model links in a natural way the low density gas probed by QSO absorption lines at high redshift and the faint-blue galaxy population seen at moderate redshifts. Third, dwarf galaxies are expected to be less clustered than larger ones (Efstathiou 1995), which at least goes in the right direction towards explaining the low amplitude of the angular correlation function at faint magnitudes. Fourth, a specific triggering mechanism (the drop in the ionizing UV background at $z \approx 1$) has been identified that would promote cooling and star formation in low-mass gas clouds at moderate redshift. Tidal interactions are assumed to provide the triggering in other dwarf-dominated models (Lacey et al. 1993; Kauffmann, Guiderdoni, & White 1994), but the details are left unspecified. Fifth, the angular size distribution of the dwarf galaxies is expected to be peaked at small scale, in qualitative agreement with recent results from the HST Medium Deep Survey (Im et al. 1995). Finally, because wholesale merging is not required in the models, normal galaxies, particularly the giant galaxies, can follow standard passive evolution tracks in color and luminosity *vs.* redshift, as favored by observations of cluster ellipticals (Aragón-Salamanca et al. 1993; Rakos & Schombert 1995; Dickinson 1995) and luminous field galaxies (Colless 1994; Steidel & Dickinson 1995).

While the model does not contradict any specific observations, there are circumstantial arguments *against* the "boojums as starbursting dwarfs" hypothesis:

1. We must postulate that the faded remnants of the bursting dwarf population, while populating the local universe at a density of $\sim 2$–$4\,\mathrm{Mpc}^{-3}$, are for the most part unobserved. Their structural properties are similar to those of the local group dwarf spheroidals, but their star formation histories are different. The local group dSph galaxies all show evidence for an old (Globular Cluster age) population, by virtue of RR-Lyrae stars and hence, argue against the hypothesis that the formation of low-mass galaxies is suppressed by the background UV flux.

   On the other hand, hierarchical clustering models also predict that low-mass density peaks that reside in regions destined to form groups and clusters will virialize at earlier epochs than those residing in the "field". As we noted in §2, typical (*i.e.* $1\sigma$) peaks on $10^9\,M_\odot$ scale in an $\Omega = 1$, $\sigma_8 = 0.67$ CDM universe, for example, virialize at $z \approx 3.5$ whereas $2\sigma$ peaks of the same mass scale virialize at $z \approx 8$. If the universe is not photoionized at the time when the high peaks form virialized structures, then the resulting halos will certainly experience starbursts soon after virialization. In addition, halos that form at higher redshifts are more compact. At $z = 8$, for instance, a virialized $10^9\,M_\odot$ halo has a circular velocity of $V_c = 38\,\mathrm{km\,s}^{-1}$; even if the universe is photoionized at this epoch, the $T \approx 3 \times 10^4$ K gas will not remain "stably confined" in such halos but will cool and eventually form stars. It is quite likely that the local group dwarf spheroidals do not correspond to "typical" peaks but rather, high $\sigma$ peaks. It should be noted that high peaks are expected to be strongly clustered. Furthermore, it should also be noted that the peculiar episodic star formation histories of the local group dSphs are problemic for any model for galaxy formation.

2. Recent evidence for metal enrichment in the Ly$-\alpha$ forest (Tytler 1995; Cowie et al. 1995) suggests that the intergalactic gas is not of primordial abundance. It is not clear, therefore, whether the metagalactic radiation field will be able to overcome the more efficient cooling. However, the detections indicate that the metal abundance in the Ly$-\alpha$ forest is $\sim 0.01 Z_\odot$ and for such low values of $Z$, the cooling function of the gas not significantly different from that for gas of primordial abundance as the cooling is dominated by hydrogen and helium recombination line radiation.

3. In our models the faded remnants of the boojum population are low-surface brightness (LSB) galaxies that have ejected all their gas. In contrast, nearby LSB galaxies outside of clusters are generally gas rich. For example, 77% of the optically-selected LSB galaxies studied by Eder et al. (1989) were detected in HI at 21-cm. On the other hand, the Eder survey, however, is almost 100% incomplete for diameters less than $40''$, the very regime where the faded boojums are expected to be found.

## 5.2. Tests of the Bursting Dwarf Hypothesis

Fortunately, the model does not stand or fall on the above arguments. There are number of observations that can provide a test of the bursting dwarf scenario, without being particularly sensitive to our specific assumptions about star formation histories, dust, metallicities, or effective radii. We outline some of these tests below.

1. *The redshift distribution at very faint magnitudes should show a peak near $z = 1$.* With most of the star formation activity occurring between redshifts $0.5 \lesssim z \lesssim 1$, the bursting dwarf model makes a unique prediction for the redshift distribution at magnitudes $B > 25$ where the dwarf galaxies dominate the counts. The exact shape of the peak in the redshift distribution shown in Figure 13 is determined by $t_*$ and our assumption that the dwarfs do not form before $z = 1$. However, even a more realistic spread in formation redshifts to $z > 1$, will not remove the pronounced peak.

2. *The K-band counts should continue to rise to $K_{AB} = 28$.* Passive evolution models with $q_0 = 0.5$ are able to match the K-band counts down to magnitudes, $K_{AB} \approx 24$, but predict continued flattening of the $N(m)$ distribution to fainter magnitudes. In contrast, the bursting dwarf model predicts that slope of the $N(m)$ distribution should stay constant or even rise fainter than $K = 24$. The near-infrared light from the bursting dwarfs counteracts the expected decline in slope. It is worth noting that the dwarfs fade in the $K$ band as well as the optical bands, and provide most of their contribution to the counts near the starburst epoch. Thus the predicted redshift distribution at $K_{AB} \approx 25$ is similar to that at $B_{AB} = 25$.

3. *The angular diameters of galaxies should continue to decrease with magnitude.* In our models, the effective radii of the dwarfs is $\sim 1$ kpc and varies by a factor of $\sim 10$ over the luminosity function. At $z = 1$, a galaxy with $r_e = 1$ kpc would have an observed $r_e = 0.1''$, near the limits of HST resolution with the WFPC camera. Because the counts in the bursting-dwarf scenario are dominated by galaxies near the minimum of the $\theta(z)$ relation, the predicted angular sizes are probably the smallest that could be anticipated in *any* plausible model of galaxy evolution with $q_0 = 0.5$.

4. *The local field luminosity function from surveys with total magnitude selection ought to exhibit a steeply rising turn-up at magnitudes fainter than $M_B \approx -16$.* The exact magnitude at which this rise begins will depend on the details of the star formation histories and whether or not this upturn is detected depends sensitively on the selection criteria of the survey. For example, the luminosity function recovered from our simulated catalogs using the isophotal selection of Loveday et al. (1992) survey shows no evidence of the steepening in spite of the large number of low luminosity dwarfs being present.

5. *Faded, low surface-brightness remnants of the boojums population should be detectable at low redshift.* The predicted redshift distribution, $K$-counts, and angular diameter distribution could perhaps be reproduced by a model involving tidally triggered star formation in satellite dwarf galaxies (Lacey et al. 1993; Kauffmann, Guiderdoni, & White 1994), if the star formation activity were suitably tuned to happen near $z = 1$. However, in such models the dwarfs are accreted onto the larger of the normal galaxies by low redshifts and few isolated remnants are expected. In contrast, our model 1 predicts between 2–4 galaxies per cubic Mpc, integrated over the luminosity function. If the galaxies expand during the supernova wind phase, as we assume, their surface brightnesses should be very low, but not beyond the detection limits of surveys currently underway, such as the driftscan survey of Dalcanton (1995). In the HST F555W band (essentially $V$-band), a rough estimate, based on the selection criteria of Dalcanton (1995), is between 12–25 LSB galaxies per square degree with central surface brightness $\mu_0 < 26$ mag arcsec$^{-2}$ and exponential scale lengths greater than $3''$, and between 5–10 galaxies per square degree with exponential scale lengths greater than $4''$. It should be noted that the detection limits for the galaxies depend on both the luminosity and surface brightness. While the luminosity estimates, which depend on the fading of the stellar populations, probably cannot be much different than what we have assumed, the surface brightness estimates are much less secure as they depend on our assumption about the expansion of the galaxy during the supernova-wind phase.

6. *There should be an appreciable rate of Type Ia supernovae associated with the faded dwarfs.* While there are large uncertainties in the detectability of the galaxies, if the binary star fraction is similar to that in normal galaxies, the presence of the faded boojums could be signaled type Ia supernovae. This would be true even if the boojums exploded entirely and left behind a sea of intergalactic stars. Because the nature and evolution of the progenitors of type Ia supernovae are uncertain (Tutukov, Yungelson, & Iben 1992; Della Valle & Livio 1994), we can make at best a rough estimate of the supernova frequency. Adopting a rate of $2.9 \times 10^{-3}$ supernovae per year per $10^{10} L_{\odot_B}$ from van den Bergh & Tammann (1991), and using the luminosity density for the dwarfs computed in §4.5, there should be $9.5 \times 10^{-6}$ SN yr$^{-1}$ Mpc$^{-3}$. For a peak absolute magnitude of $B = -19.8$, this translates to 90 SN per year brighter than $B = 15$ over the entire sky. This is sufficiently bright to be within the range of modern large-area surveys for transient events, but sufficiently faint that it is possible that such a rate would have escaped detection. For deeper searches, it is very likely that such isolated supernovae would have escaped detection. The recent photographic searches have all identified visible galaxies either before searching for SNe, or before carrying out spectroscopic follow up (Hamuy 1993; Meuler 1993; Pollas 1993; McNaught 1993). The number of supernovae expected in something like the CTIO search (Hamuy et al. 1993) is rather small in any case. That search identified 25 type Ia SNe over a period of two years on roughly 1000 Schmidt plates. If the SN rate scales with luminosity, the number of "intergalactic" supernovae expected would be 0.8, even if the search had been completely unbiased with respect to visible galaxies.

## 6. Summation

Studies of the high redshift galaxies, of the galaxies associated with QSO absorption lines and of the faint galaxies themselves suggest that the latter consists of two populations. The properties of the brighter of these faint galaxies (for example, those with $B \lesssim 24$) are consistent with their being the intermediate redshift counterparts of the locally observed population of galaxies. However, the properties of the fainter galaxies are different, suggesting that at these magnitudes, the galaxy population becomes increasingly dominated by a different class of galaxies. Motivated by the large numbers of low mass galaxies that generically arise in realistic hierarchical clustering scenarios for structure formation, various authors (*e.g.* Lacey & Silk 1991; Babul & Rees 1992; Gardner, Cowie, & Wainscoat 1993; Lacey et al. 1993; Kauffmann, Guiderdoni, & White 1994) have proposed that these very faint galaxies (boojums) are dwarf galaxies undergoing a short burst of star formation.

The main difference between the various models is the mechanism(s) for triggering star formation at the appropriate epoch.

We consider the "boojums as bursting dwarfs" model of Babul & Rees (1992), which links the epoch of formation of dwarf galaxies to the intensity of the background ionizing UV flux. This model argues that the dwarf galaxies will undergo a relatively short burst of star formation before supernova explosions expel all the gas and that thereafter, these galaxies will simply fade away. The bursting dwarf model has several attractive features, the three most important being (1) that the model is firmly grounded in the generally accepted hierarchical clustering scenarios for structure formation, (2) a large population of low mass halos is a generic feature of any hierarchical model with realistic initial conditions on galactic scales, and (3) the identification of a specific mechanism (*i.e.* the decline in the intensity of the background UV radiation from its estimated value at $z \approx 2$) that triggers starburst in these halos at moderate redshifts.

In and of itself, the model is difficult to analyze and compare to the observations as it only describes the formation and evolution of dwarf galaxies. Consequently, we augment the Babul & Rees (1992) model with a phenomenological prescription for the formation and evolution of the locally observed population of galaxies (E, S0, Sab, Sbc, and Sdm types). Ideally, one would prefer a model that details the formation and evolution of all types of galaxies from initial density fluctuations. However, the pathways leading to the formation of the observed population of galaxies are neither not well constrained nor well understood, whereas the star formation histories along the traditional Hubble sequence are well constrained by observations. We explore our hybrid model in detail using spectral-synthesis methods and Monte Carlo simulations of deep faint galaxy surveys. We find that for reasonable choices of star formation histories for the dwarf galaxies (the mass function and the size distribution of the dwarf galaxies are derived from realistic initial conditions for hierarchical clustering and are not adjusted to fit the data), the model results are in very good agreement with the observations. In particular,

1. The model number counts are consistent with those observed in $B$, $I$ and $K$ bands to the faintest magnitudes surveyed. In all three bands, the dwarfs begin to dominate at $m_{AB} \sim 24$.

2. The model redshift distributions based on galaxy selection procedures that attempt to mimic the selection criteria of individual deep redshift surveys are also in good agreement with their observationally determined counterparts. Although the model predicts a large number of relatively nearby faded/fading dwarf remnants, such galaxies are difficult to detect because of their low surface brightness.

3. At present-day as well as $z \approx 1$, the distribution of colors of the "regular" galaxies in our model are consistent with observations. Comparisons of the model color distributions of faint galaxies are also in good agreement with the observations, with colors becoming rapidly bluer in the regime dominated by the bursting dwarfs. More interestingly though, the model does not predict an overabundance of very blue galaxies in deep $K$-band surveys in spite of the fact that the dwarfs are predominantly very blue. In fact, it appears to be somewhat lacking in the blue galaxies with $K < 21$ seen by Djorgovski et al. (1995).

4. Looking back towards $z \approx 1$, the "regular" galaxies experience very modest luminosity evolution, brightening by 0.1 to 0.7 mag in rest-frame $B_J$ and -0.3 to 0.6 mag brighter in $K$-band, with respect to the present

5. At magnitudes where the model galaxy counts are dominated by the dwarf galaxies at redshifts $0.5 \lesssim z \lesssim 1$, the angular effective radii are predicted to be a few tenths of an arcsecond. Since the angular-size *vs.* redshift relation in a $q_0 = 0.5$ universe has a minimum at $z \approx 1.2$, the angular size distribution predicted by the bursting dwarf model comes close to the extreme that could be expected for any model in a spatially flat universe. Recent deep HST images reveal that the faint galaxies are very small; such observations favor dwarf-dominated models.

These comparisons to observations indicate that the bursting dwarf (or actually, the hybrid bursting dwarf) model is a viable explanation to the puzzle of the faint galaxies. In light of recent results from gravitational lensing, angular correlation and size distribution studies, the model may have some advantages over other competing explanations. Furthermore, as set forth in this paper, the model is amenable to specific observational tests. In particular, it predicts that (see §5.2) that number counts in the $K$-band should continue to rise steeply at magnitudes fainter than $K_{AB} \approx 24$–25; that the redshift distribution at $B \gtrsim 25$ should be peaked at $z \lesssim 1$, and that a large population of nearby low-surface brightness dwarf galaxies should be detectable in surveys with a low isphototal detection threshold or through searches for apparently isolated Type Ia supernovae.

The bursting dwarf model is still rather schematic, due to the uncertainties in the evolution of the UV background, the density profiles of gas in mini-halos, and the physics of star formation. Nevertheless, it appears to offer a promising solution to the combined problems of the counts, redshift distributions, sizes, and clustering properites of very faint galaxies.

We wish to thank Cedric Lacey, Simon Lilly, Mauro Giavalisco, Stefano Casertano, Mark Dickinson, Karl Glazebrook and Richard Ellis for stimulating discussions and helpful suggestions. We are particularly grateful to Simon Lilly, Mark Dickinson and Karl Glazebrook for providing us access to their observational results prior to publication, and to Alessandro Bressan for providing us with his spectral energy distributions for comparison with our own. Finally, we wish to credit Neal Katz for coining the moniker/acronym "boojums". Support for this work is provided in part by NASA grant #HF-1043, awarded by the Space Telescope Science Institute, which is operated by the Association of Universities for Research in Astronomy, Inc., for NASA under contract NAS5-26555.